\documentclass[conference]{IEEEtran}
\IEEEoverridecommandlockouts

\usepackage{cite}
\usepackage{amsmath,amssymb,amsfonts}
\usepackage{graphicx}
\usepackage{textcomp}
\usepackage{xcolor}
\usepackage{booktabs}
\usepackage{url}
\usepackage{siunitx}
\usepackage{multirow}
\usepackage{balance}

\def\BibTeX{{\rm B\kern-.05em{\sc i\kern-.025em b}\kern-.08em
    T\kern-.1667em\lower.7ex\hbox{E}\kern-.125emX}}

\begin{document}

\title{Validation of KESTREL EMT for Industrial Capacitor Switching Transient Studies: Energization, Voltage Magnification, and VFD Interaction Analysis}

\author{\IEEEauthorblockN{Shankar Ramharack}
\IEEEauthorblockA{\textit{IEEE Power and Energy Society} \\
Trinidad and Tobago, W.I.\\
sramharack@ieee.org}
\and
\IEEEauthorblockN{Rajiv Sahadeo}
\IEEEauthorblockA{Trinidad and Tobago, W.I.\\
rajivsahadeo@outlook.com}
}

\maketitle

\begin{abstract}
Electromagnetic transient (EMT) simulation is essential for analyzing sub-cycle switching phenomena in industrial power systems, yet commercial platforms present significant cost barriers for smaller utilities, consultancies, and institutions in developing regions. This paper validates KESTREL EMT, a free open-source electromagnetic transient solver with Python integration, through three progressive case studies of industrial capacitor switching transients. Case~1 validates single capacitor bank energization against closed-form analytical solutions, achieving frequency agreement within 1.2\% and peak voltage within 3.9\%. Case~2 demonstrates the voltage magnification phenomenon at a facility low-voltage bus supplied through a Dyn transformer, quantifying a 0.79$\times$ magnification factor and confirming transformer vector group influence on transient transfer. Case~3 extends the analysis to include a 6-pulse diode front-end variable frequency drive (VFD), revealing a 1.69~p.u. transient overvoltage at the utility capacitor bank and 158\% DC bus overshoot during VFD energization---phenomena invisible to steady-state analysis. Equipment stress analysis identifies negative protection margins for surge arresters ($-$35.7\%) and VFD DC bus capacitors ($-$14\%). The results demonstrate that KESTREL, augmented with appropriate circuit modeling techniques, produces results consistent with analytical predictions and established IEEE benchmarks, providing a validated, reproducible methodology for industrial EMT studies using freely available tools.
\end{abstract}

\begin{IEEEkeywords}
Capacitor switching transients, electromagnetic transient simulation, KESTREL EMT, voltage magnification, variable frequency drives, power quality, open-source simulation.
\end{IEEEkeywords}

\let\thefootnote\relax\footnotetext{Submitted to IEEE CaribCon 2026. Code and data: \url{https://github.com/shanks847/xxxxx}. Correspondence: sramharack@ieee.org}
\section{Introduction}
\label{sec:intro}

The proliferation of power electronic loads in industrial facilities has fundamentally changed power quality challenges. Variable frequency drives (VFDs) and other converter-interfaced loads exhibit complex interactions with grid disturbances that phasor-domain analysis cannot capture \cite{hensley1991, hensley1992}. Electromagnetic transient (EMT) simulation is required for analyzing sub-cycle switching transients, nonlinear device behavior, and resonance conditions \cite{greenwood1991}.

Capacitor switching transients represent one of the most common and consequential EMT phenomena. Greenwood \cite{greenwood1991} established the foundational analytical framework. Hensley et al. \cite{hensley1991, hensley1992} documented voltage magnification at low-voltage buses and VFD nuisance tripping due to DC bus overvoltage. Abedini et al. \cite{abedini2020} demonstrated EMT modeling for industrial capacitor bank transients validated against IEEE~1036. Shipp et al. \cite{shipp2011} documented transformer failures attributable to switching transients.

Commercial EMT platforms (PSCAD/EMTDC, EMTP-RV, ETAP~eMT) carry annual licensing costs of \$10,000--\$50,000 per seat, creating barriers for smaller consultancies, academic institutions in developing regions, and independent researchers \cite{paraemt2024}. Open-source initiatives including ParaEMT \cite{paraemt2024} and PowerSimulationsDynamics.jl \cite{psid2024} address this gap primarily for transmission-level studies.

KESTREL EMT \cite{kestrel_web} is a free EMT solver implementing Dommel's nodal analysis with trapezoidal integration \cite{dommel1969}, providing standard circuit elements, switching devices, nonlinear elements, and Python code block integration for custom control models at each simulation timestep.

This paper validates KESTREL through three progressive case studies: (1)~single capacitor bank energization validated against analytical solutions, (2)~voltage magnification through a Dyn transformer, and (3)~6-pulse VFD interaction with a utility capacitor bank during energization.

\section{System Topology and Parameters}
\label{sec:system}

\begin{figure}[!t]
\centering
\includegraphics[width=\columnwidth]{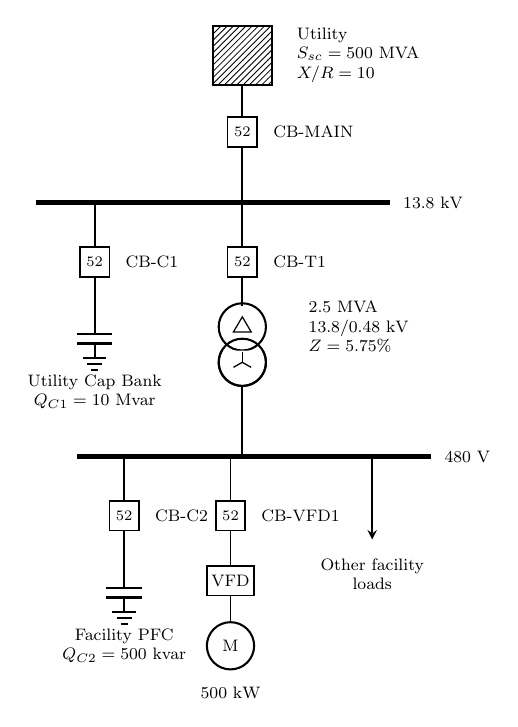}
\caption{Industrial facility single-line diagram showing utility source, capacitor bank, Dyn transformer, facility power factor correction, and VFD load.}
\label{fig:sld}
\end{figure}

The system under study represents a typical medium-voltage industrial facility supplied from a utility substation with switched capacitor banks. Fig.~\ref{fig:sld} shows the single-line diagram. Key parameters are summarized in Table~\ref{tab:system_params}.

\begin{table}[!t]
\centering
\caption{System Parameters}
\label{tab:system_params}
\begin{tabular}{lcc}
\toprule
\textbf{Parameter} & \textbf{Value} & \textbf{Unit} \\
\midrule
\multicolumn{3}{l}{\textit{Utility Source}} \\
\quad Voltage, $V_{LL}$ & 13.8 & kV \\
\quad Short-circuit capacity, $S_{sc}$ & 500 & MVA \\
\quad $X/R$ ratio & 10 & -- \\
\quad Source inductance, $L_s$ & 1.005 & mH \\
\quad Source resistance, $R_s$ & 37.9 & m$\Omega$ \\
\midrule
\multicolumn{3}{l}{\textit{Utility Capacitor Bank}} \\
\quad Reactive power, $Q_{C1}$ & 10 & Mvar \\
\quad Capacitance, $C_1$ & 139.3 & $\mu$F/phase \\
\midrule
\multicolumn{3}{l}{\textit{Transformer (Dyn, 5.75\%)}} \\
\quad Rating, $S_{xfmr}$ & 2.5 & MVA \\
\quad Secondary voltage & 480 & V \\
\quad Leakage inductance (HV ref.) & 13.95 & $\mu$H \\
\midrule
\multicolumn{3}{l}{\textit{Facility Load Bus (480~V)}} \\
\quad PFC capacitors, $Q_{C2}$ & 500 & kvar \\
\quad VFD rating, $P_{VFD}$ & 500 & kW \\
\quad DC bus nominal voltage & 650 & V \\
\quad DC bus OV trip threshold & 850 & V \\
\bottomrule
\end{tabular}
\end{table}

The 500~MVA short-circuit capacity represents a moderately stiff utility connection. The 10~Mvar capacitor bank is standard for this voltage class. The 500~kvar facility capacitor bank, at approximately 20\% of transformer rating, creates the voltage magnification condition. Source parameters are derived as follows:
\begin{equation}
Z_s = \frac{V_{LL}^2}{S_{sc}} = \frac{(13.8 \times 10^3)^2}{500 \times 10^6} = 0.381~\Omega
\label{eq:source_z}
\end{equation}

Given the $X/R = 10$, the source inductance and resistance are:
\begin{equation}
L_s = \frac{Z_s}{\omega} \cdot \frac{X/R}{\sqrt{1 + (X/R)^2}} = 1.005~\text{mH}
\label{eq:Ls}
\end{equation}
\begin{equation}
R_s = X_s / (X/R) = 37.9~\text{m}\Omega
\label{eq:Rs}
\end{equation}

The utility capacitor bank capacitance per phase is:
\begin{equation}
C_1 = \frac{Q_{C1}}{V_{LL}^2 \cdot \omega} = 139.3~\mu\text{F}
\label{eq:C1}
\end{equation}

\section{Case 1: Single Capacitor Bank Energization}
\label{sec:case1}

\subsection{Analytical Framework}

Energization of an initially uncharged capacitor bank from an inductive utility source produces a series RLC transient \cite{greenwood1991}. The natural oscillation frequency is:
\begin{equation}
f_n = \frac{1}{2\pi\sqrt{L_s C_1}} = f_{sys}\sqrt{\frac{S_{sc}}{Q_{C1}}} = 425~\text{Hz}
\label{eq:fn}
\end{equation}

Under worst-case conditions (switch closure at voltage peak, zero initial charge), the theoretical peak transient voltage approaches 2.0~p.u. \cite{greenwood1991, ieee_c37012}. The peak inrush current is:
\begin{equation}
I_{peak} = V_m\sqrt{C_1/L_s} = 4.19~\text{kA}
\label{eq:ipeak}
\end{equation}

The damping ratio $\zeta = (R_s/2)\sqrt{C_1/L_s} = 0.051$ confirms a highly underdamped system.

\subsection{KESTREL Implementation}

The KESTREL model comprises a three-phase cosine voltage source (13.8~kV line-to-line RMS), series inductance (1.005~mH/phase), series resistance (37.9~m$\Omega$/phase), a time-controlled switch closing at $t = 4.167$~ms (Phase~A voltage peak), and a wye-connected capacitor bank (139.3~$\mu$F/phase). Simulation parameters were set to a 2~$\mu$s timestep over 200~ms duration.

\subsection{Results and Validation}

\begin{figure}[!t]
\centering
\includegraphics[width=\columnwidth]{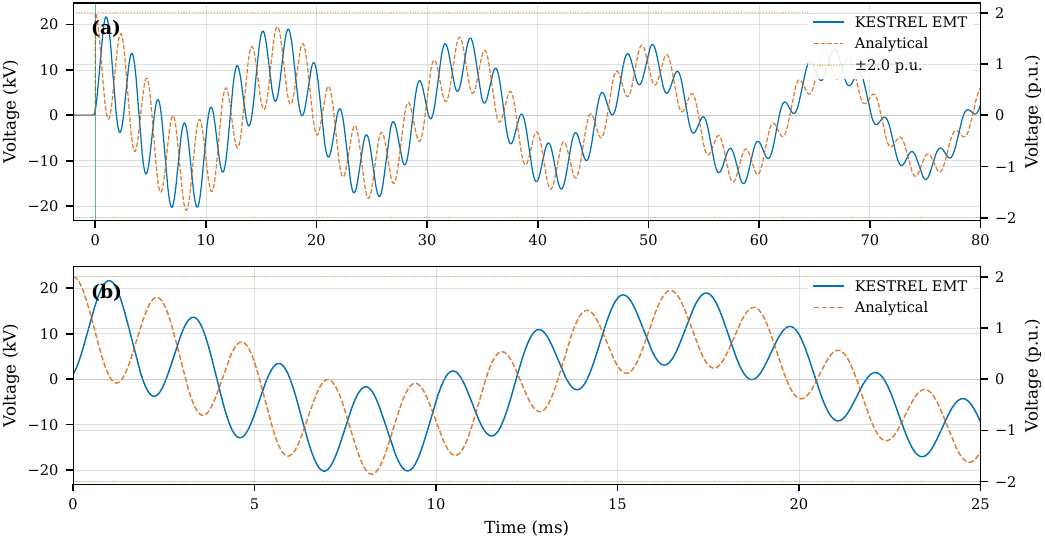}
\caption{Case~1 validation: (a) Full voltage waveform with KESTREL vs. analytical solution over 120~ms. (b)~Zoomed view of first 30~ms highlighting 425~Hz oscillation. Dashed lines indicate $\pm$2.0~p.u. theoretical limits.}
\label{fig:case1_validation}
\end{figure}

Fig.~\ref{fig:case1_validation} presents the KESTREL results compared against the analytical solution. Panel~(a) shows the full capacitor voltage waveform over 120~ms demonstrating the characteristic decaying oscillation superimposed on the 60~Hz fundamental. Panel~(b) provides a zoomed view of the first 30~ms where the 425~Hz transient oscillation is clearly visible.

Table~\ref{tab:case1_validation} summarizes the quantitative validation. The oscillation frequency measured via FFT analysis agrees within 1.2\% of the analytical prediction. Peak voltage shows 3.9\% deviation from the theoretical 2.0~p.u. maximum, attributable to finite damping. Peak inrush current agrees within 8.7\%, well within typical engineering tolerances.

\begin{table}[!t]
\renewcommand{\arraystretch}{1.2}
\centering
\caption{Case~1 Validation Results}
\label{tab:case1_validation}
\begin{tabular}{lccc}
\toprule
\textbf{Parameter} & \textbf{Analytical} & \textbf{KESTREL} & \textbf{Error} \\
\midrule
Peak voltage (p.u.) & 2.00 & 1.92 & $-$3.9\% \\
Oscillation freq. (Hz) & 425 & 420 & $-$1.2\% \\
Peak inrush current (kA) & 4.19 & 4.56 & +8.7\% \\
\bottomrule
\end{tabular}
\end{table}

\begin{figure}[!t]
\centering
\includegraphics[width=0.85\columnwidth]{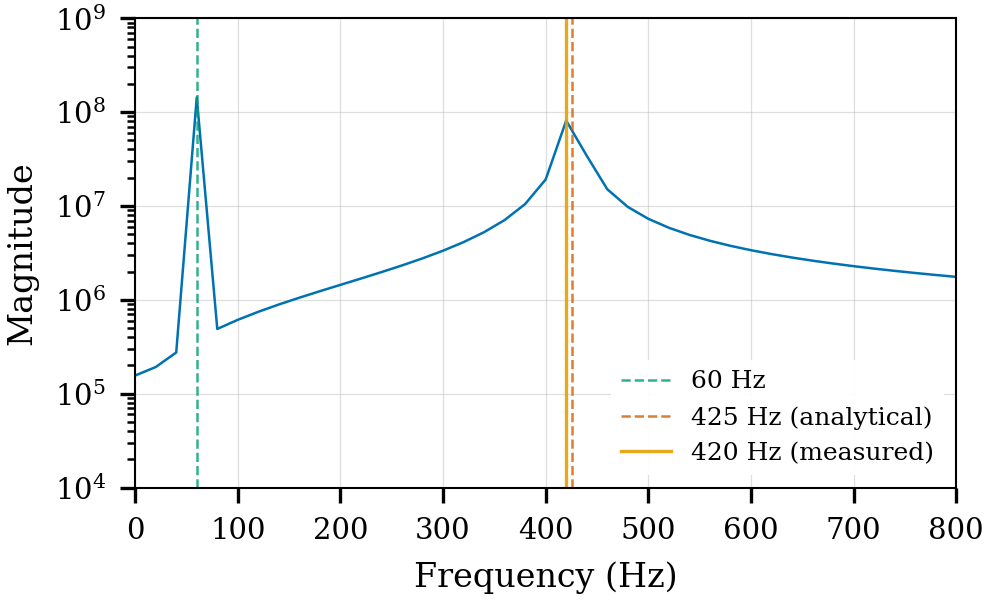}
\caption{FFT spectrum of Case~1 transient voltage showing the 60~Hz fundamental and 420~Hz oscillation.}
\label{fig:case1_fft}
\end{figure}

The FFT spectrum (Fig.~\ref{fig:case1_fft}) confirms both the 60~Hz fundamental and 420~Hz transient oscillation, closely matching the analytical prediction. The 1.2\% frequency error validates correct LC resonance implementation. Peak voltage below 2.0~p.u. is consistent with finite damping ($\zeta = 0.051$), and 8.7\% current overshoot reflects sensitivity to the precise switching instant \cite{abedini2020}.

\section{Case 2: Voltage Magnification}
\label{sec:case2}

\subsection{Phenomenon Description}

Voltage magnification occurs when a transient oscillation from utility capacitor switching couples through a distribution transformer and excites resonance at a facility's LV bus \cite{mcgranaghan1992, bhargava1993}. When the facility has PFC capacitors, the transformer leakage inductance and facility capacitance form a resonant circuit. Maximum magnification (up to 4$\times$) occurs when the tuning ratio $f_{utility}/f_{facility}$ approaches unity \cite{dugan2012}.

\subsection{Resonance Analysis}

From Case~1, $f_{utility} = 425$~Hz. The facility-side resonant frequency is:
\begin{equation}
f_{facility} = \frac{1}{2\pi\sqrt{L_{xfmr} \cdot C_{facility}}} = 562~\text{Hz}
\label{eq:ffac}
\end{equation}

The tuning ratio of 0.76 indicates frequencies sufficiently close for significant voltage magnification. However, the Dyn transformer configuration provides inherent attenuation by trapping zero-sequence transient components in the delta primary winding.

\subsection{Results}

\begin{figure}[!t]
\centering
\includegraphics[width=\columnwidth]{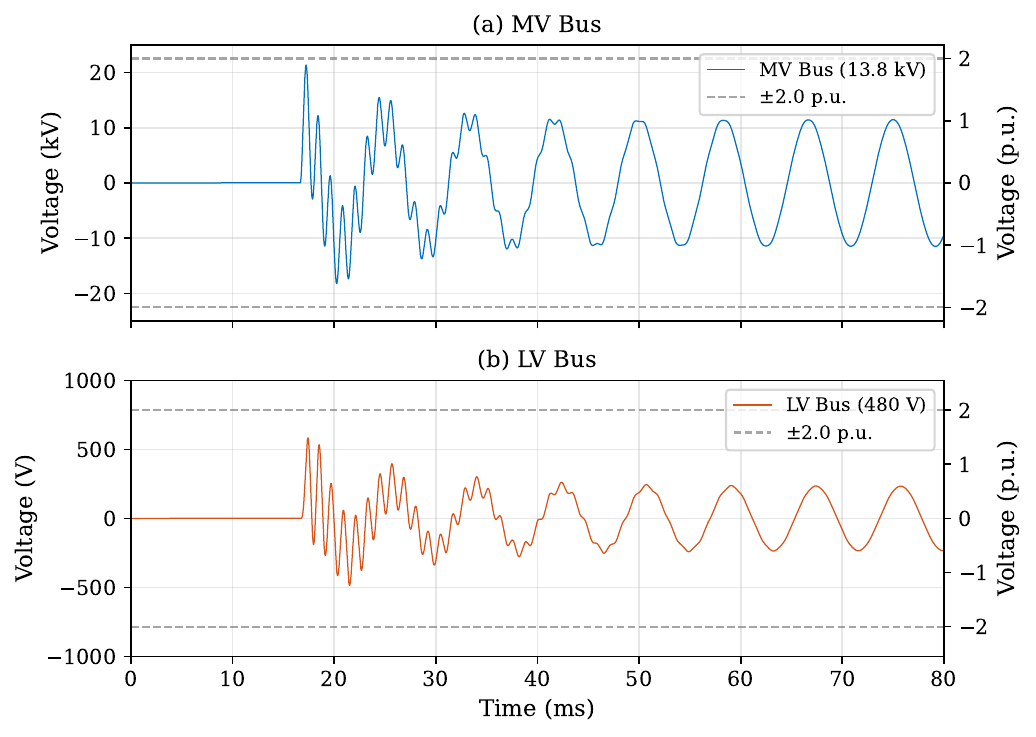}
\caption{Case~2 voltage waveforms: MV bus at utility capacitor bank and LV bus at facility secondary.}
\label{fig:case2_voltage}
\end{figure}

\begin{figure}[!t]
\centering
\includegraphics[width=0.85\columnwidth]{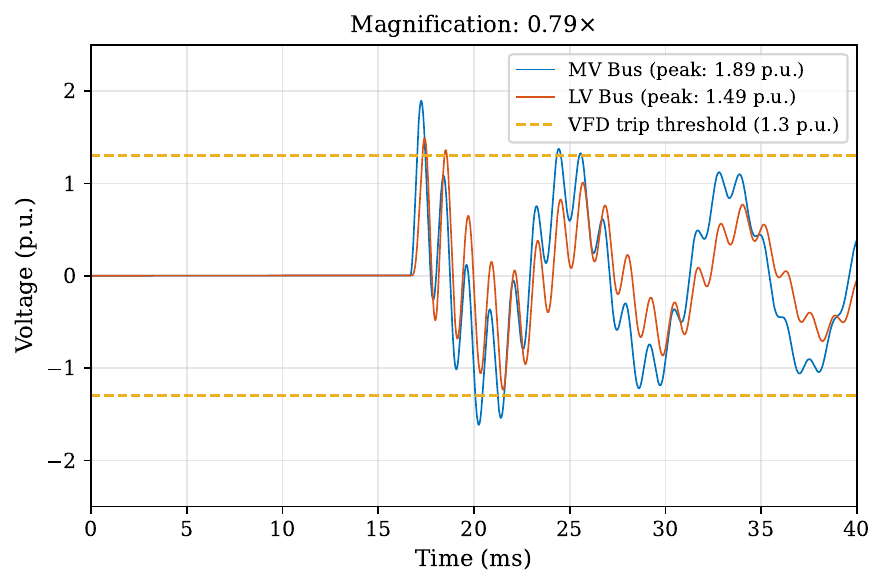}
\caption{Case~2 per-unit voltage magnification comparison. The Dyn transformer yields a magnification factor of 0.79$\times$.}
\label{fig:case2_magnification}
\end{figure}

Table~\ref{tab:case2_results} summarizes the voltage magnification results. Fig.~\ref{fig:case2_voltage} shows the MV and LV bus voltage waveforms, and Fig.~\ref{fig:case2_magnification} presents the per-unit magnification comparison.

\begin{table}[!t]
\renewcommand{\arraystretch}{1.2}
\centering
\caption{Case~2: Voltage Magnification Results}
\label{tab:case2_results}
\begin{tabular}{lcc}
\toprule
\textbf{Parameter} & \textbf{MV Bus} & \textbf{LV Bus} \\
\midrule
Peak transient & 21.35 kV & 584 V \\
Peak transient (p.u.) & 1.89 & 1.49 \\
Dominant frequency & 400 Hz & 400 Hz \\
\midrule
\textbf{Magnification factor} & \multicolumn{2}{c}{\textbf{0.79$\times$}} \\
\bottomrule
\end{tabular}
\end{table}

The 0.79$\times$ magnification factor demonstrates that the Dyn transformer provides inherent attenuation by trapping zero-sequence transient components in the delta primary \cite{martinez2005}. Despite this, the LV peak of 1.49~p.u. still exceeds the typical VFD DC bus OV trip threshold of 1.3~p.u. \cite{hensley1991}, validating transformer vector group selection as a passive mitigation strategy while highlighting its insufficiency alone to prevent VFD tripping.

\section{Case 3: VFD Interaction with Utility Capacitor Bank}
\label{sec:case3}

\subsection{Background}

VFD rectifier front-ends generate harmonic currents that interact with system impedances, particularly capacitor banks \cite{ieee519, arrillaga2003}. The 6-pulse rectifier's characteristic harmonics ($h = 6n \pm 1$) can excite parallel resonances \cite{grady2012}, and Dugan et al. \cite{dugan2012} reported industrial capacitor bank failures traced to such harmonic resonance.

\subsection{System Description and Resonance Analysis}

\begin{figure}[!t]
\centering
\includegraphics[width=\columnwidth]{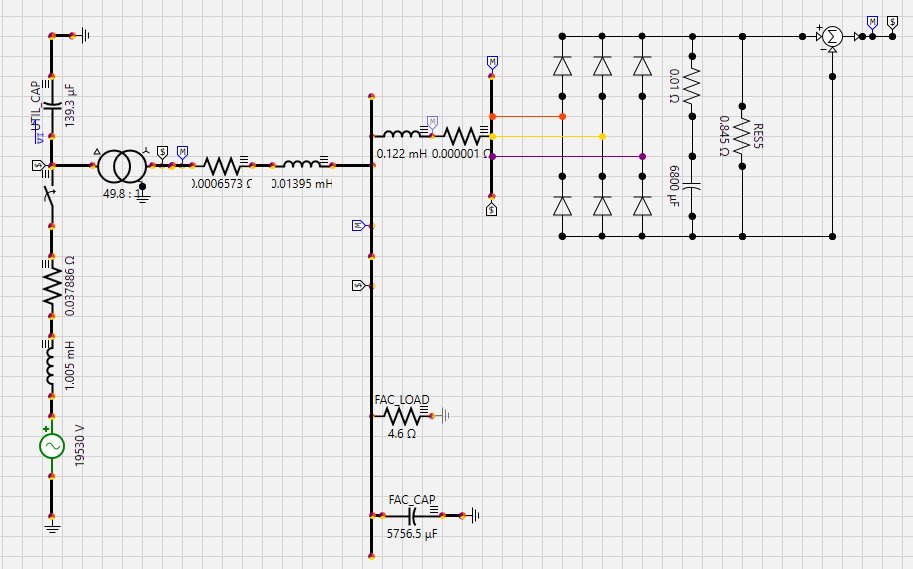}
\caption{Case~3 KESTREL circuit: 13.8~kV/480~V industrial facility with utility capacitor bank and 6-pulse VFD.}
\label{fig:case3_circuit}
\end{figure}

The system extends Case~2 with a 500~kW 6-pulse diode rectifier VFD (DC link capacitor 6,800~$\mu$F, DC load $R_{dc} = 0.845$~$\Omega$), filter inductance $L_{flt} = 0.122$~mH, facility capacitor $C_{fac} = 5{,}756.5$~$\mu$F, and facility load $R_{fac} = 4.6$~$\Omega$. Fig.~\ref{fig:case3_circuit} shows the KESTREL circuit.

The system exhibits two resonant frequencies. The HV-side parallel resonance is:
\begin{equation}
f_{res,HV} = \frac{1}{2\pi\sqrt{(L_s + L_{xfmr}) \cdot C_{util}}} = 425~\text{Hz}
\label{eq:fres_hv}
\end{equation}

This corresponds to harmonic order $h = 425/60 = 7.08$, falling precisely at the 7th harmonic---a characteristic current harmonic of the 6-pulse rectifier ($h = 6n \pm 1$) \cite{mohan2003}. The LV-side resonance is:
\begin{equation}
f_{res,LV} = \frac{1}{2\pi\sqrt{L_{flt} \cdot C_{fac}}} = 190~\text{Hz}
\label{eq:fres_lv}
\end{equation}

This places the LV resonance near the 3rd harmonic ($h = 3.17$), which can amplify triplen harmonics from unbalanced conditions. The voltage magnification at resonance is governed by the quality factor $Q = (1/R_s)\sqrt{L_s/C_{util}} \approx 71$, indicating potential for significant amplification during transient excitation \cite{xu2005, das2015}.

\subsection{KESTREL Implementation Notes}

For the Dyn transformer model, the turns ratio relates primary L-N peak to secondary L-N peak voltage: $V_{source} = 277\sqrt{2} \times 49.8 = 19{,}530$~V (L-N peak). The DC bus voltage must be measured differentially between positive and negative rails using a sum block due to the floating DC bus reference.

\subsection{Results}

\subsubsection{Transient Overvoltage}

\begin{figure}[!t]
\centering
\includegraphics[width=\columnwidth]{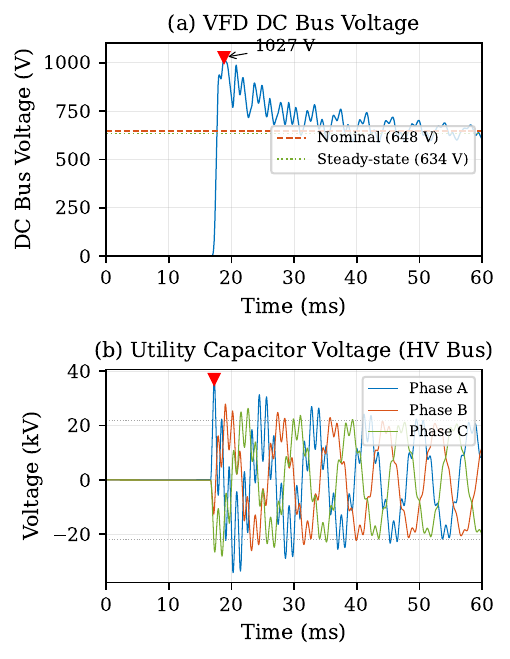}
\caption{Case~3 transient response: (a)~VFD DC bus voltage showing inrush and settling. (b)~Utility capacitor voltage showing 1.69~p.u. transient overvoltage.}
\label{fig:case3_transient}
\end{figure}

Fig.~\ref{fig:case3_transient} shows the system response during VFD startup. The utility capacitor bank experiences 1.69~p.u. transient overvoltage on Phase~A, exceeding the typical 1.4~p.u. temporary overvoltage (TOV) capability specified for metal-oxide surge arresters per IEEE C62.22 \cite{ieee_c62_22}. The DC bus exhibits a significant inrush transient reaching 1,027~V (158\% overshoot) before settling to steady-state.

Table~\ref{tab:case3_overvoltage} summarizes the peak overvoltage by phase.

\begin{table}[!t]
\renewcommand{\arraystretch}{1.2}
\centering
\caption{Case~3 Transient Overvoltage by Phase}
\label{tab:case3_overvoltage}
\begin{tabular}{lccc}
\toprule
\textbf{Phase} & \textbf{Peak (kV)} & \textbf{p.u.} & \textbf{Time (ms)} \\
\midrule
A & 37.1 & 1.69 & $\sim$8 \\
B & 27.9 & 1.27 & $\sim$12 \\
C & 28.0 & 1.28 & $\sim$15 \\
\bottomrule
\end{tabular}
\end{table}

\subsubsection{DC Bus and Harmonic Analysis}

\begin{figure}[!t]
\centering
\includegraphics[width=\columnwidth]{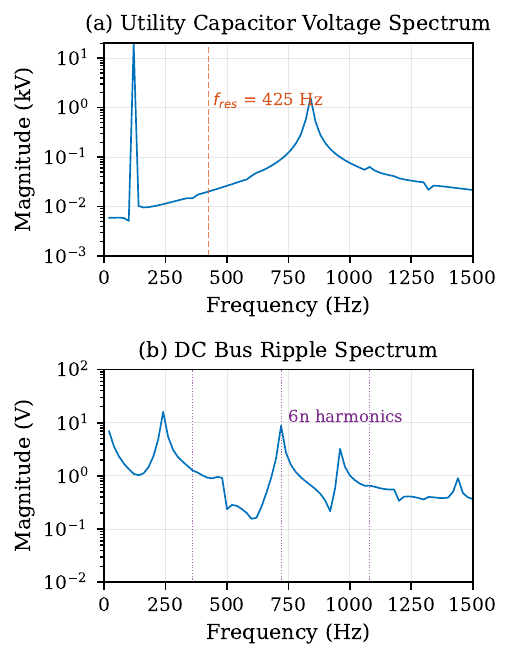}
\caption{Case~3 frequency analysis: (a)~Utility capacitor voltage spectrum showing resonance at 425~Hz. (b)~DC bus ripple spectrum confirming 720~Hz dominance.}
\label{fig:case3_fft}
\end{figure}

Steady-state DC bus characteristics: mean voltage 633.7~V (97.8\% of ideal 648~V), ripple 100.2~V peak-to-peak (15.8\%), dominant ripple at 720~Hz (12th harmonic). The FFT analysis (Fig.~\ref{fig:case3_fft}) confirms the 425~Hz system resonance in the capacitor voltage spectrum and 720~Hz dominance in DC bus ripple, consistent with $6n$ characteristic voltage harmonics \cite{mohan2003}.

\subsubsection{Validation Against Analytical Predictions}

Table~\ref{tab:case3_validation} compares analytical predictions with KESTREL results, demonstrating close agreement across all parameters.

\begin{table}[!t]
\renewcommand{\arraystretch}{1.2}
\centering
\caption{Case~3 Analytical vs. Simulated Results}
\label{tab:case3_validation}
\begin{tabular}{lcc}
\toprule
\textbf{Parameter} & \textbf{Analytical} & \textbf{KESTREL} \\
\midrule
HV resonant frequency & 425 Hz & Confirmed \\
LV resonant frequency & 190 Hz & Confirmed \\
DC bus voltage & 648 V (ideal) & 634 V (97.8\%) \\
DC ripple frequency & 720 Hz & 720 Hz \\
Transient overvoltage & $>$1.5 p.u. & 1.69 p.u. \\
VFD power & 500 kW & 475 kW \\
\bottomrule
\end{tabular}
\end{table}

\subsection{Equipment Stress Analysis}

The transient phenomena have significant implications for equipment ratings and protection coordination. Table~\ref{tab:case3_stress} summarizes the equipment stress analysis.

\begin{table}[!t]
\renewcommand{\arraystretch}{1.2}
\centering
\caption{Case~3 Equipment Stress Analysis}
\label{tab:case3_stress}
\begin{tabular}{lccc}
\toprule
\textbf{Equipment} & \textbf{Rating} & \textbf{Stress} & \textbf{Margin} \\
\midrule
\multicolumn{4}{l}{\textit{Surge Arrester (24~kV class)}} \\
TOV (0.1~s) & 27.3 kV & 37.1 kV & \textbf{$-$35.7\%} \\
Energy (Class 2) & 58.5 kJ & $<$1 kJ & +98\% \\
\midrule
\multicolumn{4}{l}{\textit{Capacitor Bank (IEEE Std 18~\cite{ieee_std_18})}} \\
Voltage (110\%) & 15.6 kV & 37.1 kV & \textbf{$-$138\%} \\
Current (135\%) & 1,003 A & 887 A & +12\% \\
\midrule
\multicolumn{4}{l}{\textit{VFD DC Bus}} \\
OV trip threshold & 842 V & 1,027 V & \textbf{$-$22\%} \\
DC capacitor & 900 V & 1,027 V & \textbf{$-$14\%} \\
\bottomrule
\end{tabular}
\end{table}

The 37.1~kV transient exceeds the surge arrester TOV capability by 35.7\% (for a 24~kV arrester, $V_{TOV,0.1s} = 1.40 \times 19.5 = 27.3$~kV per IEEE C62.22 \cite{ieee_c62_22}). Repeated events cause cumulative thermal stress and potential failure. The capacitor bank peak represents 262\% of rated voltage, yielding $(2.62)^2 = 6.9\times$ normal dielectric stress per IEEE Std~18 \cite{ieee_std_18}. The VFD DC bus peak of 1,027~V exceeds both the OV trip threshold (842~V) and typical capacitor rating (900~V) \cite{epri2000}.

These transient phenomena occur faster than conventional protection can respond: arrester TOV damage in milliseconds versus fuse clearing at 10--100~ms; DC bus overvoltage in microseconds versus contactor opening at 20--50~ms \cite{ieee_c37_99}. This underscores the need for preventive measures (detuning reactors, pre-insertion resistors, active front-end drives) rather than reactive protection \cite{das2015}.

\section{Discussion}
\label{sec:discussion}

The progressive case study approach demonstrates KESTREL's capability across increasing complexity. A key finding from Case~3 is the ``steady-state blind spot'': conventional load flow or harmonic frequency scan would report normal voltages and improved power factor with no indication of the 1.69~p.u. transient overvoltage---this phenomenon requires time-domain EMT simulation \cite{dugan2012, das2015}.

Practical modeling considerations include KESTREL's voltage source scaling convention (``VLL~RMS''), the Dyn transformer turns ratio specification on a per-phase peak-voltage basis, and the differential DC bus measurement requirement for floating rectifier topologies. These details, while KESTREL-specific, highlight the general importance of model verification before conducting EMT studies \cite{martinez2005}.

The validation errors (1.2\% frequency, 3.9\% voltage, 8.7\% current) are well within engineering tolerances. The current overshoot reflects sensitivity to the precise switching instant, a known characteristic of capacitor energization \cite{greenwood1991, abedini2020}.

\section{Conclusion}
\label{sec:conclusion}

This paper validated KESTREL EMT for industrial capacitor switching transient studies through three progressive case studies. Key contributions include:

\begin{enumerate}
\item Quantitative validation of KESTREL against closed-form analytical solutions with frequency agreement within 1.2\% and peak voltage within 3.9\% for fundamental capacitor energization transients.

\item Demonstration of voltage magnification through a Dyn transformer configuration, quantifying a 0.79$\times$ magnification factor and confirming transformer vector group influence as a passive mitigation strategy.

\item Identification of a 1.69~p.u. transient overvoltage and 158\% DC bus overshoot during VFD energization---phenomena invisible to steady-state analysis---with equipment stress analysis revealing negative protection margins for surge arresters, capacitor banks, and VFD DC components.

\item Documentation of practical KESTREL implementation details including voltage source conventions, transformer turns ratio specification, and differential DC bus measurement, establishing reproducible methodologies for practitioners.
\end{enumerate}

The results demonstrate that KESTREL EMT, as a freely available tool, produces results consistent with analytical predictions and established IEEE benchmarks for industrial capacitor switching studies. This contributes to the democratization of EMT simulation capability, particularly for practitioners, consultancies, and institutions in developing regions where commercial platform licensing costs present prohibitive barriers.

Future work will extend this validation to include active front-end VFD response, battery energy storage system transient support, surge arrester coordination studies, and comparison with commercial EMT platform results.

\section*{Acknowledgment}
The authors thank David T. Daigle for developing and freely distributing KESTREL EMT, and for responsive technical support during model development.

\balance
\bibliographystyle{IEEEtran}
\bibliography{references}

\end{document}